\documentclass[prb,twocolumn,url,nofootinbib]{revtex4}
\usepackage{amsmath,bm}
\usepackage{graphicx,pdflscape}% Include figure files
\usepackage{color}
\usepackage[toc,page]{appendix}

\usepackage{hyperref}
\usepackage{color,subfigure}
\usepackage{placeins}

\newcommand{\beq}{\begin{equation}}
\newcommand{\eeq}{\end{equation}}
\newcommand{\barray}{\begin{eqnarray}}
\newcommand{\earray}{\end{eqnarray}}

\newcommand{\refdisp}[1]{Ref.~[\onlinecite{#1}]}
\newcommand{\figdisp}[1]{Fig.~(\ref{#1})}

\newcommand{\si}{\sigma}

\renewcommand{\Re}{\mathrm{Re}}
\renewcommand{\Im}{\mathrm{Im}}

\renewcommand{\emph}{\textit}

\newcommand{\X}[2]{X_{{#1}}^{#2}}

\newcommand{\chem}{\bm {\mu}}

\newcommand{\ket}{\rangle}

\usepackage{atveryend}
\makeatletter
\let\origcitation\citation
\AtEndDocument{\def\mycites{}%
  \def\citation#1{\g@addto@macro\mycites{#1^^J}\origcitation{#1}}}
\AtVeryEndDocument{\newwrite\citeout\immediate\openout\citeout=\jobname.cit
  \immediate\write\citeout{\mycites}\immediate\closeout\citeout}
\makeatother

\begin{document}

\title{The $t$-$t'$-$J$ model in one dimension using extremely correlated Fermi liquid theory \\ and time dependent density matrix renormalization group}
%\title{Momentum  dependence of self-energy in the one dimensional $t$-$t'$-$J$ model}
\author{Peizhi Mai$^1$, Steven R. White$^2$ and B. Sriram Shastry$^1$\\
$^1$Physics Department, University of California, Santa Cruz, CA 95064 \\
$^2$Department of Physics and Astronomy, University of California, Irvine, CA 92717 \\
}
\date{\today}

\begin{abstract}
We study  the one-dimensional  $t$-$t'$-$J$ model for generic couplings using two complementary theories, the  {\em  extremely correlated Fermi liquid theory} and  {\em time dependent density matrix renormalization group} over a broad energy scale.  The two methods 
 provide a unique  insight into the strong  {\em momentum dependence} of the  self-energy 
 of this prototypical non-Fermi liquid, described at low energies as a  Tomonaga-Luttinger liquid. We also demonstrate  its  intimate relationship to spin-charge separation, i.e. the splitting of  Landau quasi-particles of higher dimensions into two constituents, driven by  strong  quantum fluctuations inherent in one dimension. The momentum distribution function, the spectral function, and the excitation dispersion of these two methods also compare well.
\end{abstract}

\maketitle

\section{\bf Introduction} In varying  dimensions the  $t$-$J$model continues to
attract  attention owing to its relevance in cuprates and other important
strongly interacting electronic systems. The model  embodies very strong
correlations, which    lie outside the regime of validity of perturbation
theory, and thus pose a  challenging problem. Our main goal in this work is to obtain an understanding of the properties in
one dimension (1-d), {\em over a wide energy range}. 

At  low energies  
  the bosonization technique  has been widely  applied to the
(closely related) Hubbard model \cite{Giamarchi,1dhubbard,bosonization,Meden,CI}.
For large U several non-perturbative methods have been devised  to study the
$t$-$J$model for general dimensions,  including the study  of finite  clusters
\cite{ed,Prelovsek}  and large-N based slave particle mean-field theories
\cite{Slave-boson}. In 1-d we also have exact results using Bethe's ansatz
\cite{BA1,BA2,BA3,BA4,BA5,BA6}  at special values of the parameters of the
model, and also  for long-ranged versions \cite{LRtJ} of the $t$-$J$model, using
techniques developed in  the Haldane-Shastry models. Photoemission experiments \cite{Dardel} have been carried
out to study the spectral properties of several quasi 1-d metals, relevant to the $t$-$J$model.

To study  a wider energy range,  including the  low to  intermediate and high energy regimes,  we employ and compare the results from two complementary techniques.
In 1-d, the density matrix renormalization group (DMRG) \cite{DMRG}
provides nearly exact results for the ground state, and can also be used
for finite temperature and spectral properties.
Ground state DMRG has been used to give the phase diagram of the $t$-$J$model over a broad range of parameters
in \cite{DMRG2}.
Here we study dynamics using the time dependent density
matrix renormalization group (tDMRG). tDMRG \cite{DMRG,tDMRG} has been used to
obtain virtually exact spectral functions for spin chains, but only a few times
for doped Fermi systems. One such time was a tDMRG treatment of the $t$-$J$model, obtaining
spectral functions for the system at finite temperature \cite{DMRG1}.
In this work we use tDMRG only at $T=0$, but we have pushed much farther in terms of system size,
accuracy, and frequency resolution than in \cite{DMRG1}. This accuracy is needed to resolve the detailed
features of the self-energy, which has not been done before with tDMRG.

The other technique used is  the extremely correlated
Fermi liquid  (ECFL) theory \cite{ECFL}.  This analytical theory, which can treat a large
class of large $U$ problems, including the $t$-$J$model, uses Schwinger's functional
differential equations for the electron Green's function. These equations are
systematically expanded in a  parameter $\lambda \in [0,1]$, representing
partial Gutzwiller projection. The ${\cal O}(\lambda^2)$ theory leads to a
closed set of coupled equations \cite{ECFL,Pathintegrals} for the Green's
function. This treatment has been benchmarked in high dimensions and in 2-d. 
In infinite dimensions, dynamical mean field theory (DMFT) \cite{infinited}
provides  a solution to the Hubbard  model, and ECFL has been
benchmarked recently \cite{Sriram-Edward,WXD} against
exact results from the single impurity Anderson model, and DMFT in $d=\infty$
\cite{badmetal,HFL}.  The limiting case $U=\infty$  has been explored in detail in \cite{ECFL-DMFT}.  
The agreement at low energies is  good enough  to yield
accurate results for the low T resistivity,  a  highly sensitive variable.
In 2-d, ECFL has been applied recently to cuprate superconductors \cite{SP,PS}.
It is therefore interesting  to
see how well this scheme deals with the physics of 1-d. The equations used here have the character of a skeleton graph  series. We have checked that the second order skeleton graphs for the Hubbard model in 1-d already displays characteristics of spin-charge separation and non-Fermi liquid spectral functions, while the non-skeleton, i.e. bare perturbation theory  does not.

 Understanding the extent of  {\em momentum dependence} of the Dysonian
 self-energy $\Sigma$ in various dimensions is  one of the goals of the present
 work. While the $d=\infty$ models have a  momentum {\em independent}
 self-energy,  momentum dependence of $\Sigma$  is inevitable  in lower
 dimensions. However there is  a scarcity of reliable information on its extent
 and location. In most published work, the self-energy in 1-d is rarely
 presented \cite{Veljko-1d}, or even calculated, since standard  solutions
 directly deal with the Green's function.  In contrast we focus on unraveling
 the $(\vec{k},\omega)$ dependence of the Dysonian self-energy in 1-d and
 comparing with its higher dimensional counterparts.

 \begin{figure*}[t]
\subfigure[\;\; n=0.7, t'=0, J=0.3]{\includegraphics[width=.7\columnwidth]{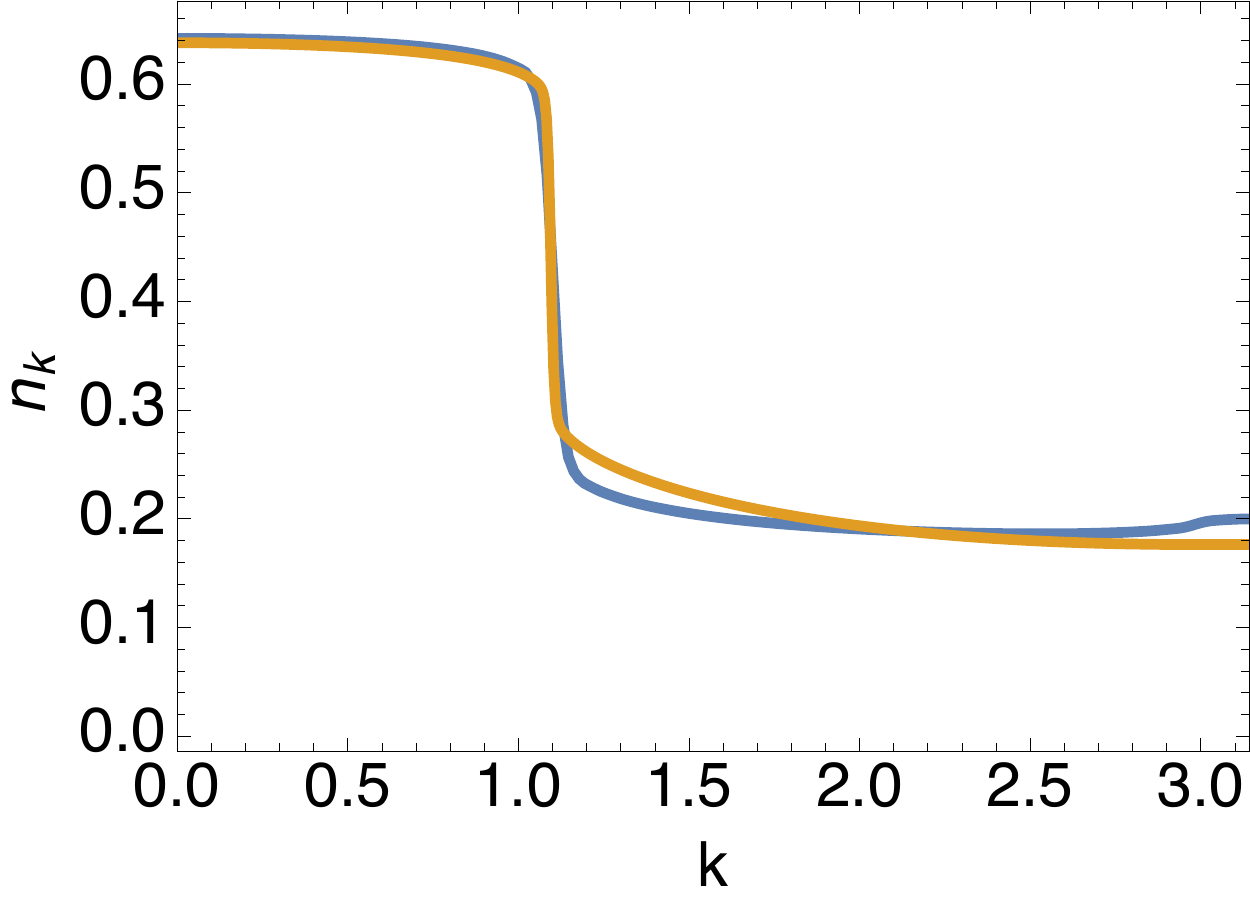}}
\subfigure[\;\; n=0.7, t'=0, J=0.6]{\includegraphics[width=.7\columnwidth]{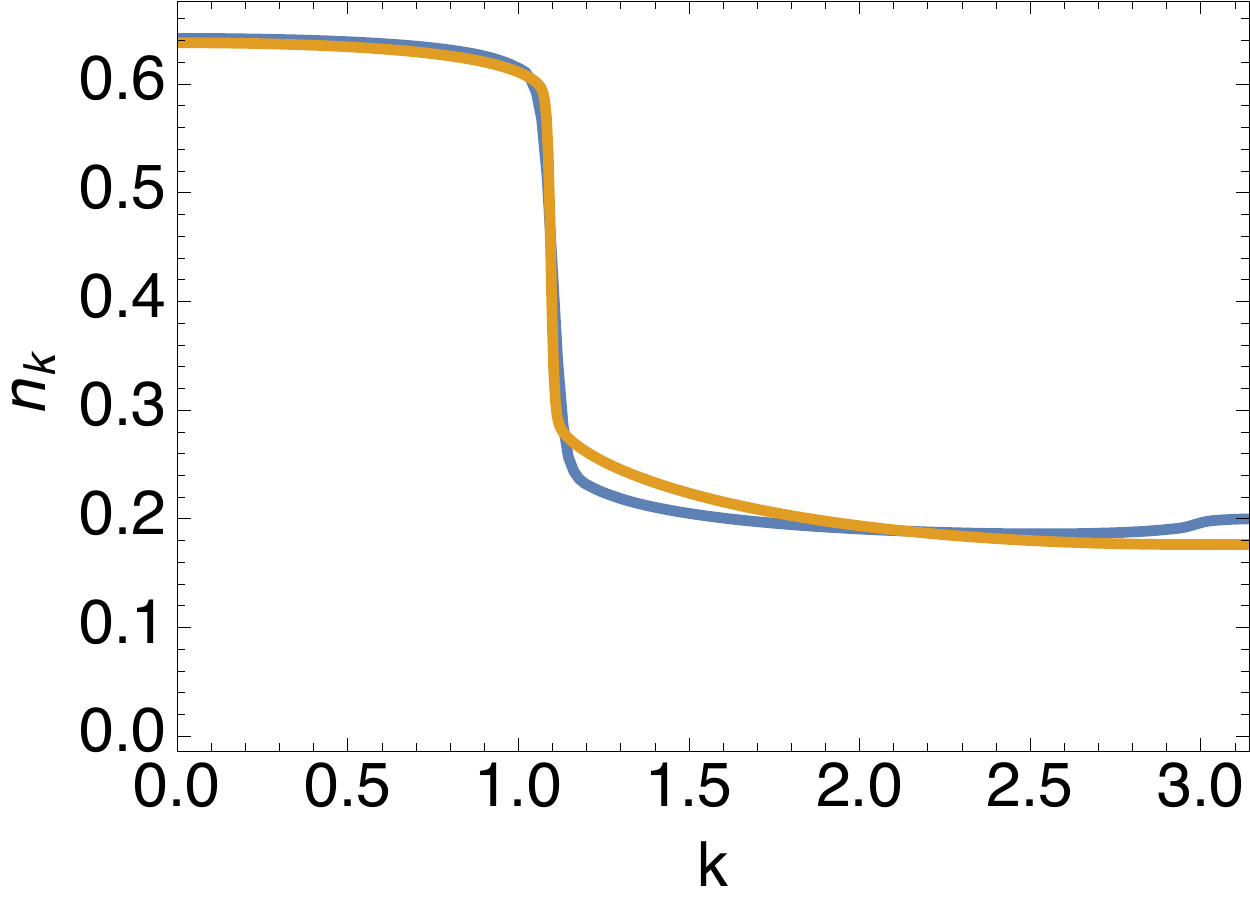}}
\subfigure[\;\; n=0.7, t'=0.2, J=0.3]{\includegraphics[width=.7\columnwidth]{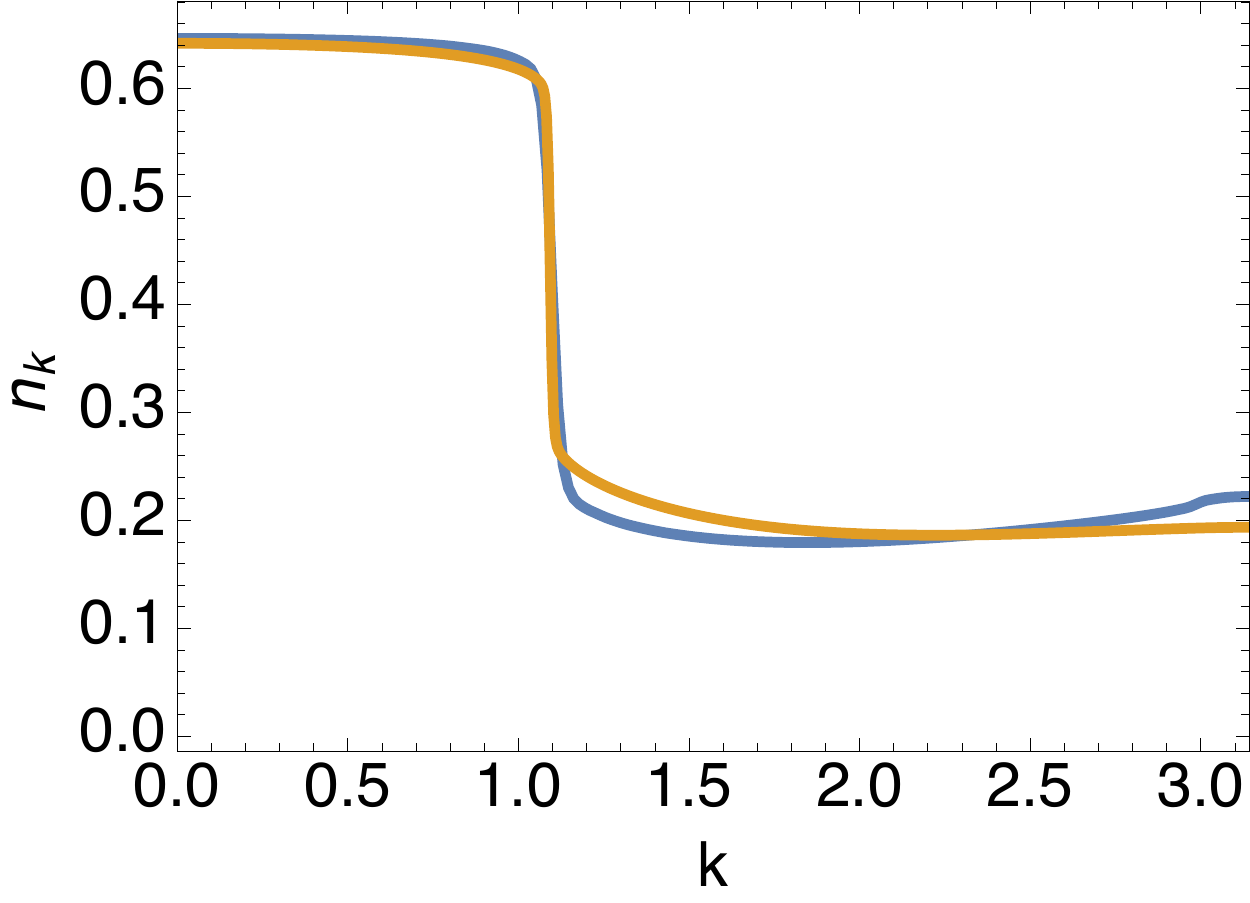}}
\subfigure[\;\; n=0.7, t'=0.2, J=0.6]{\includegraphics[width=.7\columnwidth]{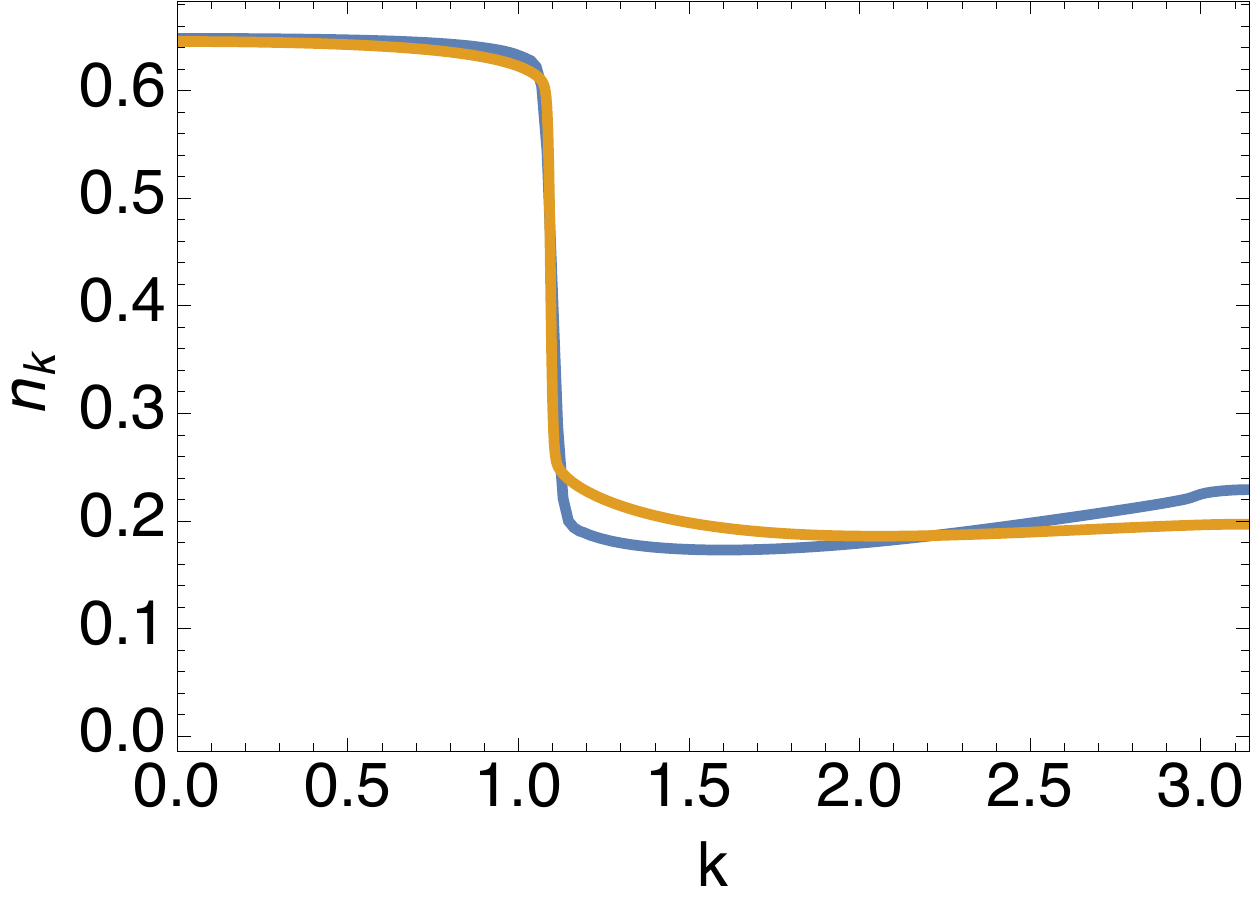}}
\caption{ \label{nk} Momentum distribution $n_k$ for  ECFL (yellow) at T=0.005 and tDMRG (blue) at T=0 with n=0.7, J=0.3, 0.6 and t'=0, 0.2. In all cases these two methods agree well especially in the occupied region and both give a power law singularity at $k_F$. The small discrepancy in the unoccupied region corresponds to the  $3k_F$ feature in the exact solutions discussed in \cite{BA1}. This subtle singularity is missed by  the ${\cal O}(\lambda^2)$ equations. 
%from both methods. This agreement indicates that ECFL captures the correct $k$, $t'$ and $J$ dependence.
}
\end{figure*}
%\FloatBarrier

\section{\bf Overview}
In the present work we solve the $d=1$ $t$-$t'$-$J$ model for generic parameters using {\em the same set of ECFL equations} as in higher dimensions.  We calculate  from the two theories the  momentum distribution function, self-energy, spectral function and excitation dispersion over a broad energy scale. 
 %In the low $k,\omega$ regime exhibiting non-Fermi liquid behavior, the agreement between the two and the  Tomonaga-Luttinger liquid (TLL) theory velocities of spinons and holons\cite{ed}, as well as anomalous exponent\cite{DMRG2}, and find reasonable agreement. 
 In the low $k,\omega$ regime exhibiting non-Fermi liquid behavior, reasonable agreement is found between the two and the exact diagonalization (ED) data in the velocities of spinons and holons \cite{ed}, as well as the Tomonaga-Luttinger liquid (TLL) theory in anomalous exponent \cite{DMRG2}.
 Extending the ${\cal O}(\lambda^2)$ ECFL equations to higher orders holds promise of a better agreement.   At higher energies, where few  studies exist,  the agreement between the two theories  is quite good already. A valuable insight  gained  at low energies is the close relationship between a momentum dependent ridge in the $\Im \, \Sigma(k,\omega)$ and the spin-charge separation.

\section{\bf Model and Parameters used}
The  Hamiltonian of the 1-d $t$-$t'$-$J$ model is
\barray
\begin{split}
H_{tJ} &= - t \sum_{\langle ij\rangle} \X{i}{\si 0}\X{j}{0\si} - t' \sum_{\langle\langle ij\rangle\rangle} \X{i}{\si 0}\X{j}{0\si} - \chem \sum_i \X{i}{\si \si}, 
\\&\ + J \sum_{\langle ij\rangle} \left( \vec{S}_i . \vec{S}_j - \frac{1}{4} \X{i}{\si \si} \X{j}{\si' \si'}   \right), \label{hamiltonian}
\end{split}
\earray
where repeated spin indices are summed , $\X{i}{\sigma 0}= P_G C^\dagger_{i \sigma} P_G$, $\X{i}{0 \sigma }= P_G C_{i \sigma} P_G$, $\X{i}{\sigma \sigma'}=  P_G C^\dagger_{i \sigma} C_{i \sigma'} P_G$ with $P_G= \Pi_i (1-n_{i \uparrow} n_{i \downarrow})$ as the Gutzwiller projection operator. $\langle ij\rangle$ and $\langle\langle ij\rangle\rangle$ refers to summing over first and second neighbor pairs respectively.

For this model \cite{ECFL,SP} 
we compute the results from the two theories at density $n=0.7$, 
%two values of
second nearest neighbor hopping 
$t'/t=0,0.2$ and 
%two values of 
$J/t=0.3,0.6$.
We avoid the special cases of $t'=0=J$ since this leads to a degenerate
spectrum, with a charge sector that is isomorphic to  the spinless Fermi gas. 
The ECFL results are shown at various $T$ while the tDMRG  results are at $T=0$
where most reliable calculations are possible. $t=1$ is the energy unit and will be neglected below.

\begin{figure*}[t]
\centering
\subfigure[\;\; ECFL, T=0.005]{\includegraphics[width=.99\columnwidth]{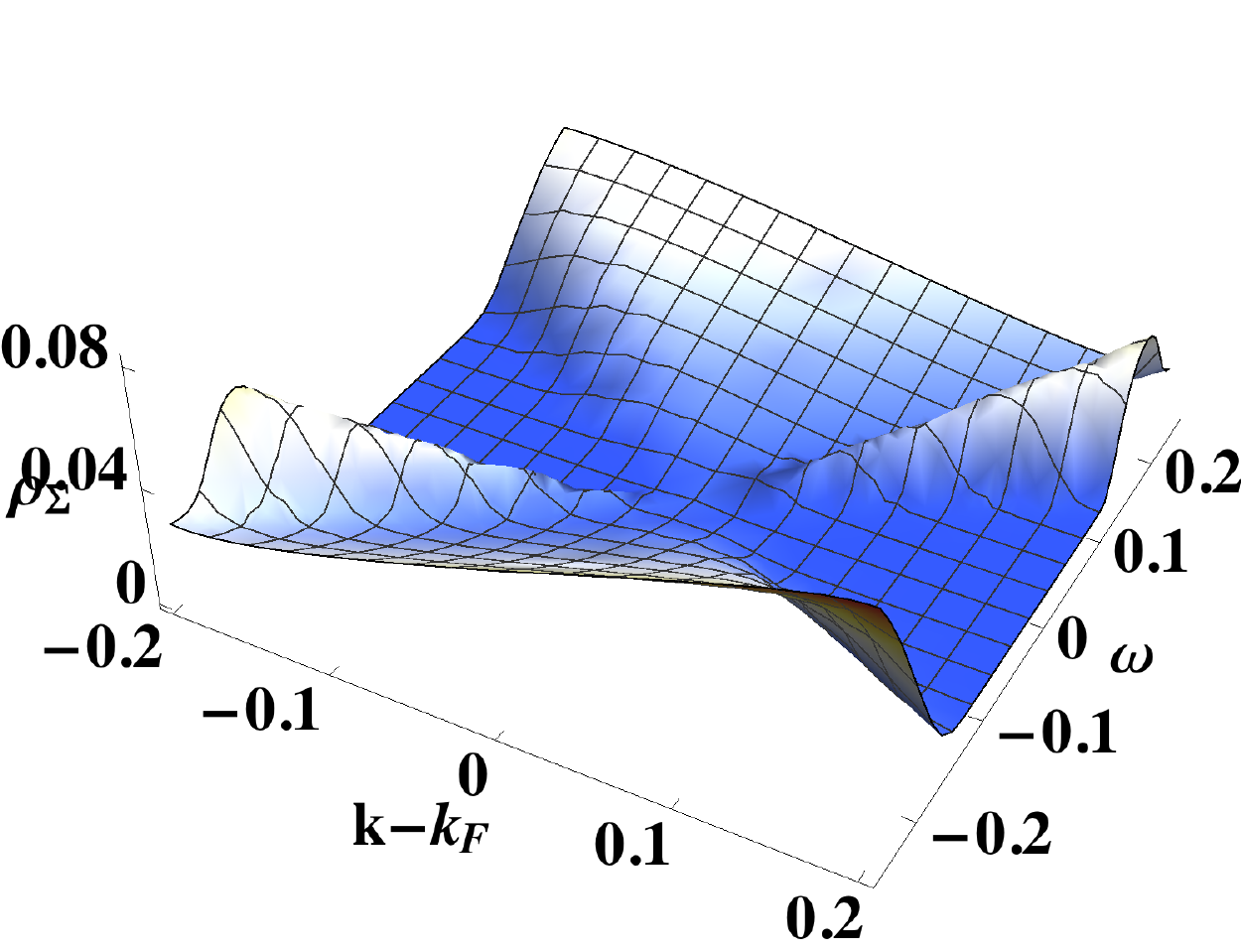}}
\subfigure[\;\; tDMRG, T=0]{\includegraphics[width=.99\columnwidth]{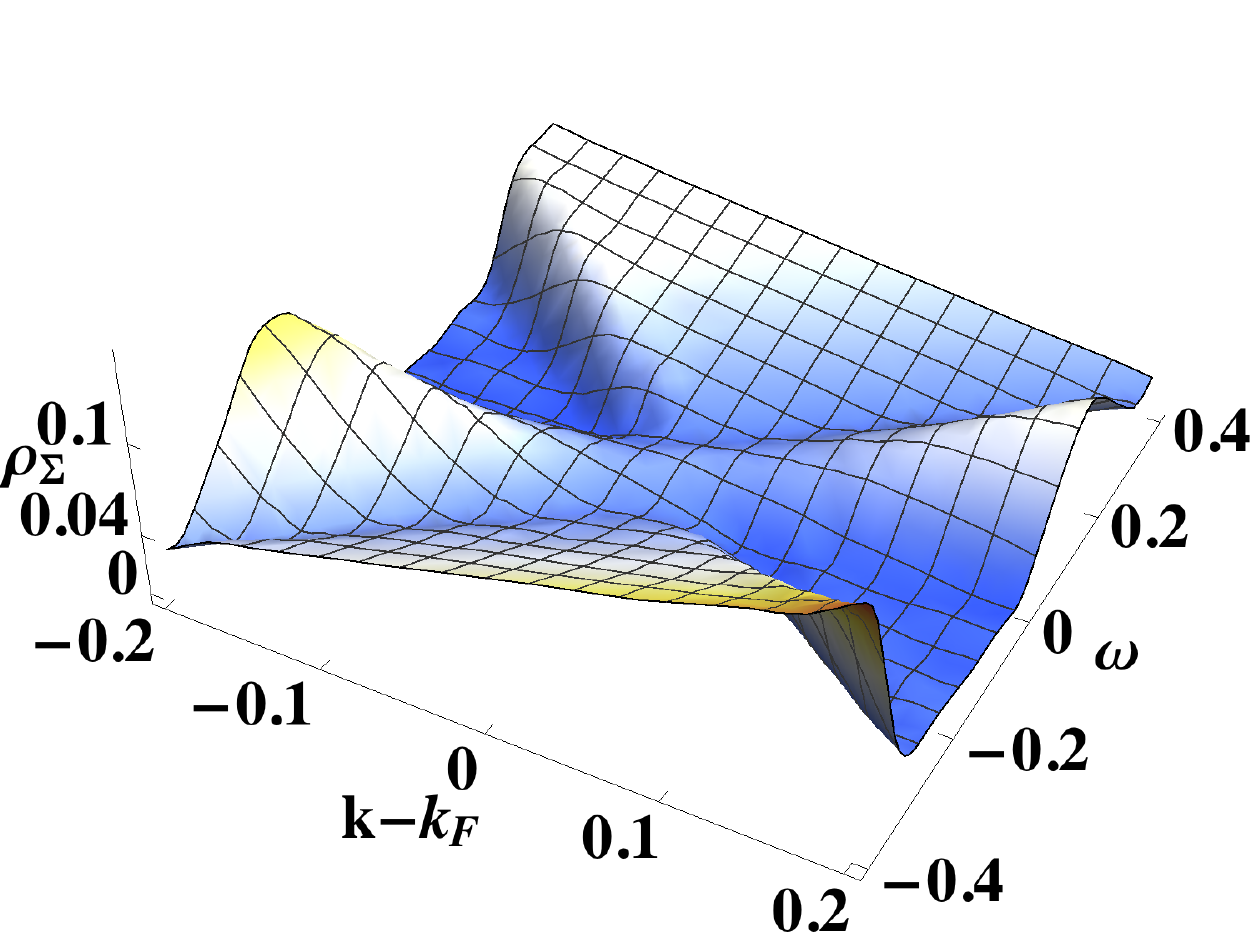}}
\caption{ \label{sigma3D}  n=0.7, J=0.3, t'=0: Imaginary  self-energy  $\rho_{\Sigma}(k,\omega)$ at  low $\omega$ and $k-k_F$ from both methods.  Both give a dominant $(k, \omega)$ dependent ridge running from left to right, and a less prominent feature  running from top-left to bottom-right. Both of them pass through $k=k_F, \omega=0$ region. The dominant ridge is responsible for the appearance of the twin peaks structure in the spectral functions which represents the spin-charge separation. The peaks for $k<k_F, \omega<0$ are seen in the left half of the electronic spectral function in \figdisp{EDC} panels  (a,b), while the peaks for $k>k_F,\omega>0$ are seen in the right half of the same figures. As seen in \figdisp{fig5} panel (c), the peak in the self-energy $\rho_\Sigma$ directly leads to a dip in the electronic spectral function $\rho_G$, provided the real part is small.  
%This type of ridge structure in momentum dependence is therefore a signature of Tomonaga-Luttinger liquids in one dimension. 
}
\end{figure*}
\begin{figure}[htbp]
\subfigure[\;\; ECFL, T=0.005]{\includegraphics[width=.49\columnwidth]{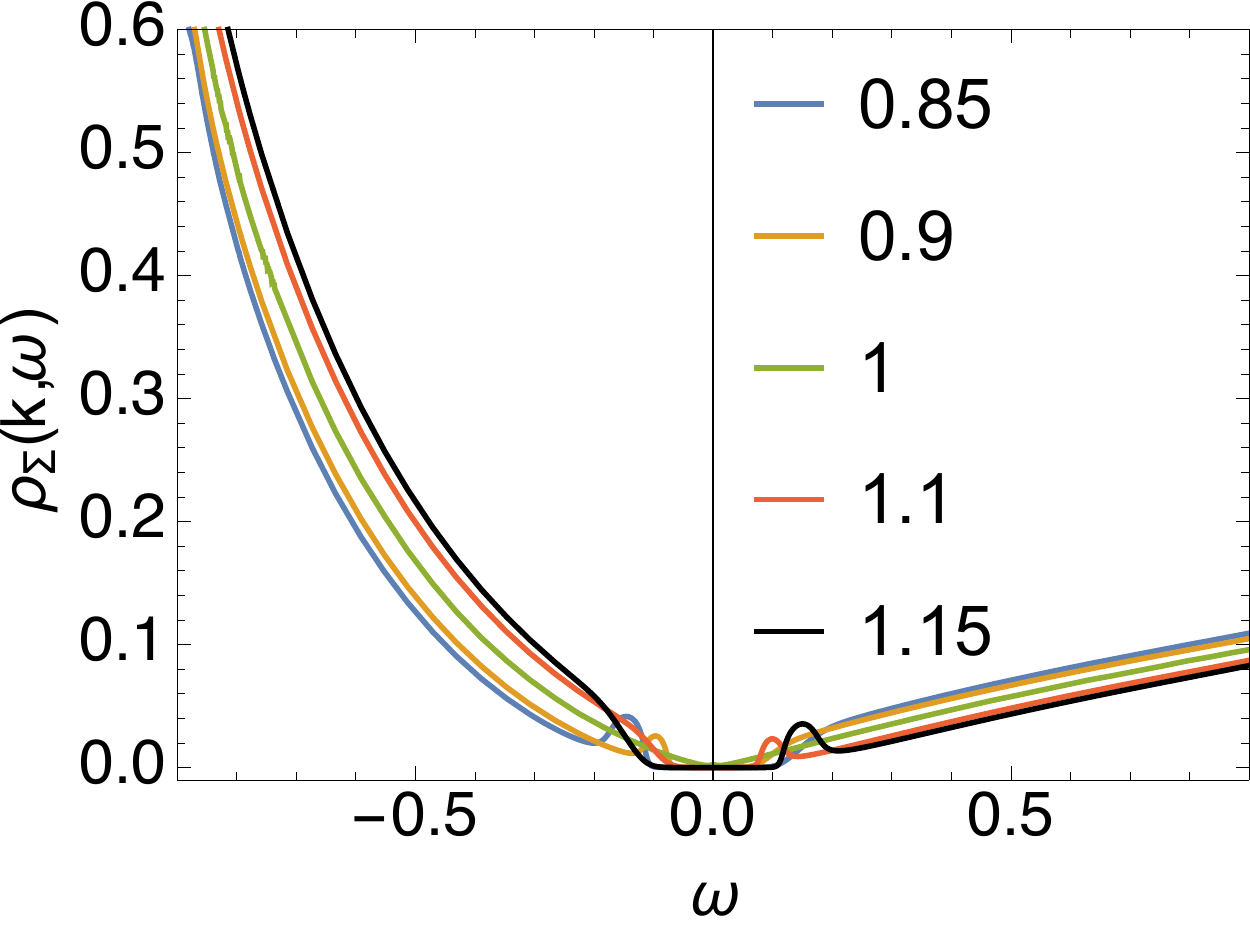}}
\subfigure[\;\; tDMRG, T=0]{\includegraphics[width=.49\columnwidth]{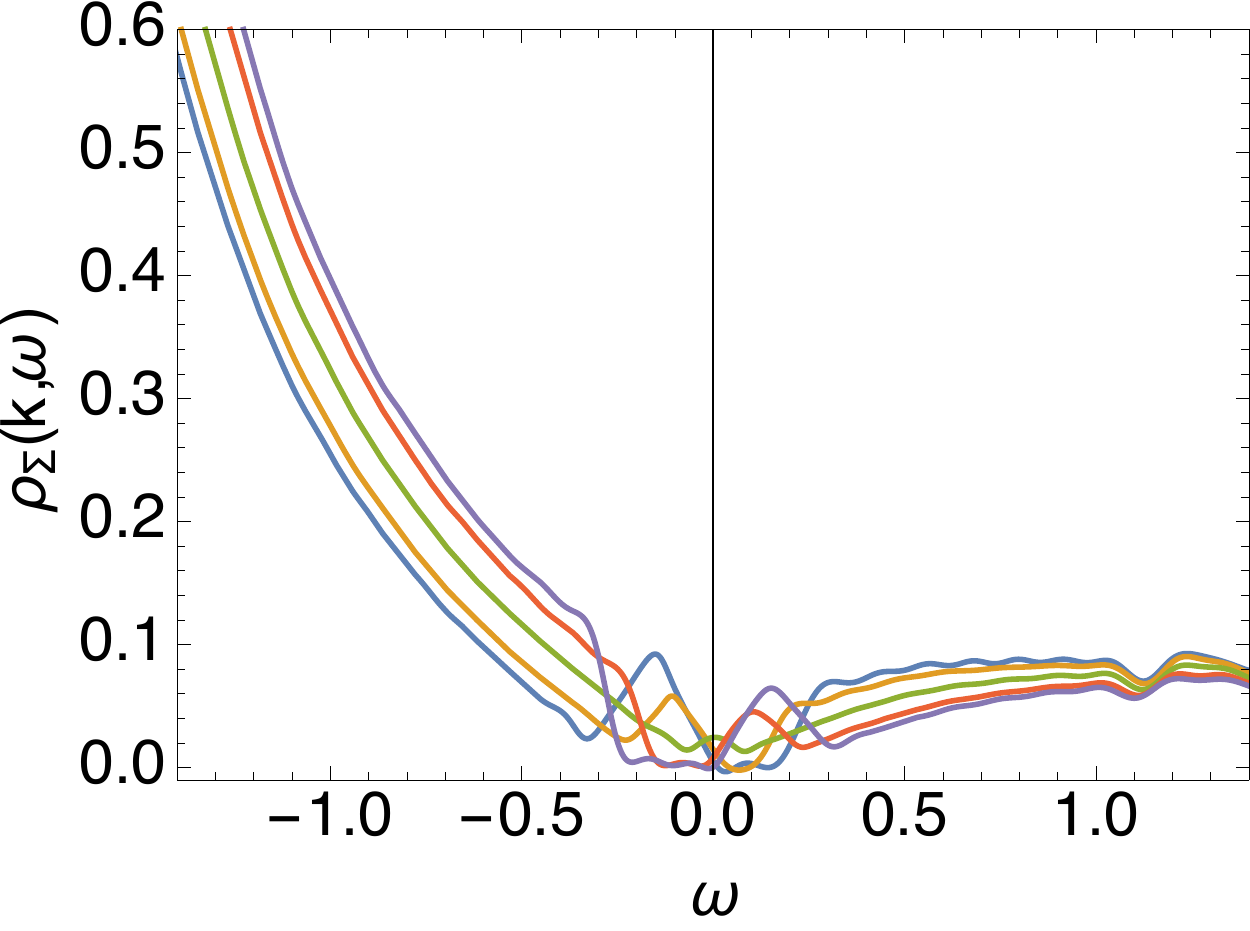}}
\caption{ \label{sigma}  $n=0.7, J=0.3$: $\rho_{\Sigma}(k,\omega)$ vs $\omega$ at   marked $k/k_F$'s.
, from ECFL at $T=0.005$ (a) and tDMRG at $T=0$ (b) in a large scale. 
The two sets of results are similar on a broad energy scale, and are comparable to higher dimensional results. 
The  low energy behavior is  discussed below. 
}
\end{figure}

{The tDMRG methods used are very similar to those used in 
\refdisp{WSKPRL}. We start by obtaining the ground state $|0\ket$ using DMRG on
a rather long but finite chain, with $L=400$, and then apply $\hat c_0$ or
$\hat c^\dagger_0$ to a site 0 near the center, forming $|\psi(t=0)\ket$. We
use a Trotter based time evolution algorithm, with
fermionic swap gates to handle next-nearest neighbor terms. We specify a density matrix eigenvalue
truncation cutoff of $3\times10^{-8}$ during the evolution, subject to a constraint on the maximum number
of states kept of $m=3000$. (Results were checked by comparing to $m=2000$.)
We evolve out to a time $t=50$.  At $t=50$, the normalization of $|\psi(t)\ket$ had decreased by
a few percent, a small error affecting primarily the widths of any sharp peaks. 
The space and time dependent Green's function is obtained by sandwiching $\hat c_i$ or $\hat c^\dagger_i$
between the ground state and $|\psi(t)\ket$ for all $i$. Linear prediction is used
to extend the time dependent Green's function out to $t=100$, after which the data is windowed 
and Fourier transformed.%From the Green's function, along with the spectral weight function, we obtain the self-energy. 
This calculation 
represents the most accurate and detailed study to date of the spectral
properties of the model at $T=0$.}

\section{\bf Momentum distribution function}
{%$n_k$ is an important quantity to study
In 1-d $t$-$J$model, $n_k$ shows a power law singularity at $k_F$ \cite{1dhubbard, CI}, a signature of the TLL, unlike a jump in higher dimensions as Fermi liquid behavior. This feature is observed from both methods in \figdisp{nk} for different $t'$ and $J$. Due to the second order approximation, the weak $3k_F$ singularity related to shadow band \cite{BA1,BA4} is not observed in ECFL results. Besides this weak effect, $n_k$ from both methods agrees well, especially in the occupied side, showing that ECFL describes the correct $t'$ and $J$ dependent behaviors.}

\begin{figure*}
\begin{minipage}[c][10cm][t]{.5\textwidth}
  \vspace*{\fill}
  \centering
  \includegraphics[width=10cm,height=7.5cm]{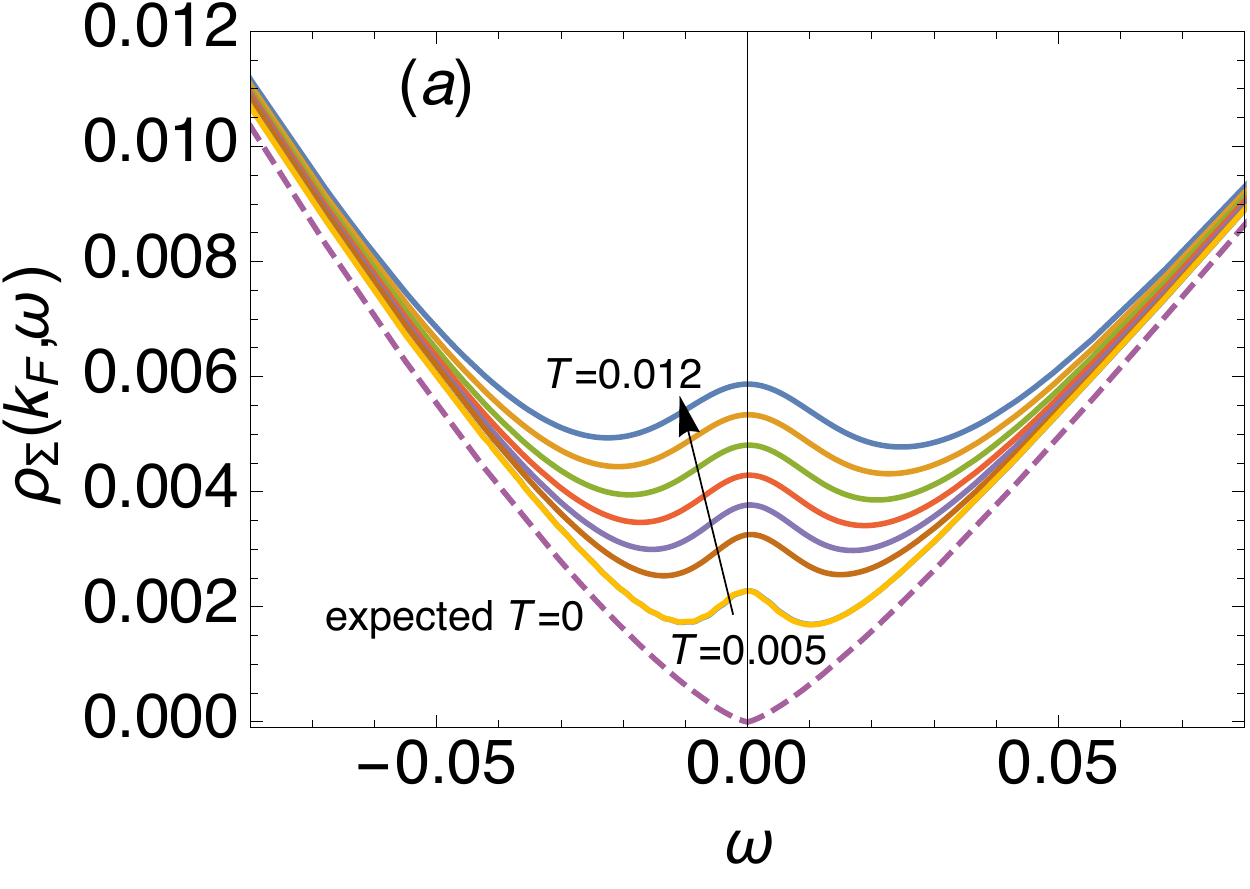}
\end{minipage}%
\begin{minipage}[c][4.5cm][t]{.45\textwidth}
  \vspace*{\fill}
  \centering
  \includegraphics[width=5cm,height=3.5cm]{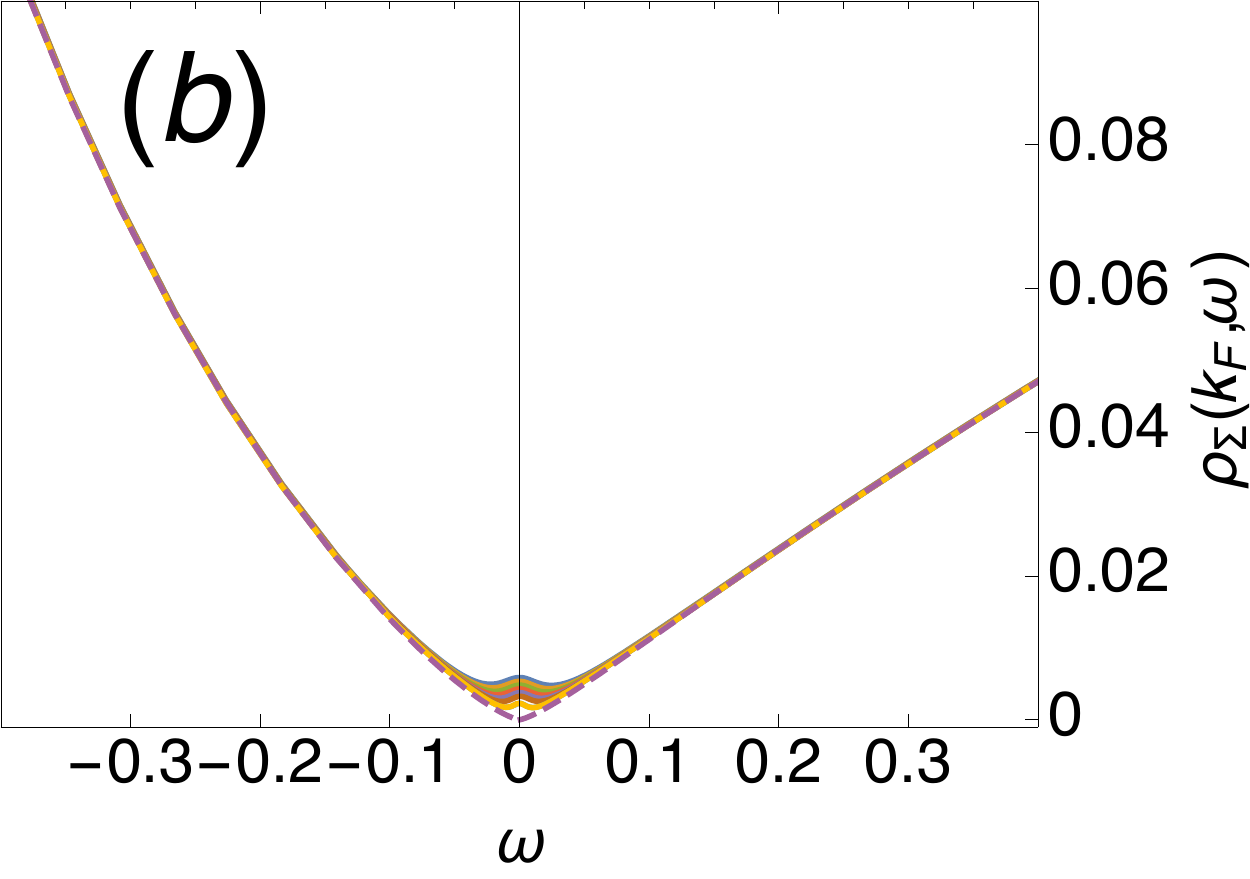}
  \label{}\par\vfill
  \includegraphics[width=5cm,height=3.5cm]{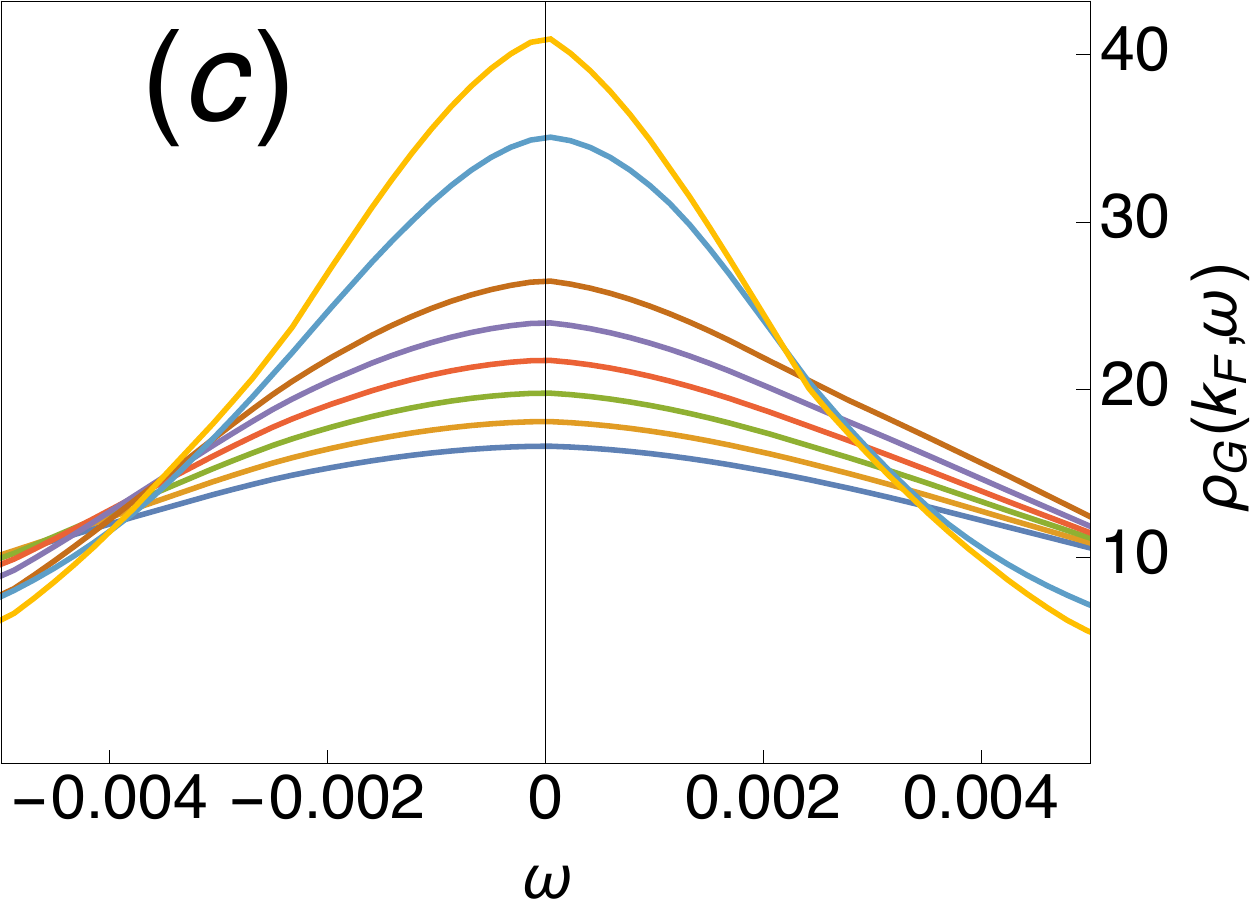}
  \label{}
\end{minipage}
\caption{ \label{sigmaTscaling}   $\rho_{\Sigma}(k_F,\omega)$ from ECFL is shown in (a) for several T at $J=0.3, t'=0$. The central peak $\rho_{\Sigma}(k_F,0)$ scales as $T^{1.1}$, in contrast to   Fermi liquid behavior $T^2$.
%with $\alpha\approx1.1$
Extrapolating to $T=0$ the double minimum structure disappears, leaving behind a $\sim|\omega|^{1.3}$ dependence.
% with $\gamma \sim 1.3$. 
(b) displays the self-energy in larger scale where changing $T$ barely makes a difference.
(c) shows the 
%peak of 
spectral function %$\rho_G(k_F,0)=1/(\pi^2\rho_{\Sigma}(k_F,0))$ 
softened by warming. }
\end{figure*}

\section{\bf Self-energy}
Next we present the Dysonian self-energy in terms of  its spectral function $\rho_{\Sigma}$ defined as
\beq
\rho_{\Sigma}(k,\omega)=-\frac{1}{\pi}\Im \, \Sigma(k,\omega).
\eeq
It is derived separately from the Green's functions in ECFL and tDMRG methods. In tDMRG,  $\Sigma$ can be found  from $G$ by inverting the Dyson relation $G^{-1}= G_0^{-1}- \Sigma$.   The ECFL theory produces two (non Dysonian) self energies $\Phi, \Psi$ \cite{ECFL}, and the resulting G can again be inverted to find the standard Dysonian $\Sigma$. Both ECFL $(T=0.005)$ and tDMRG $(T=0)$ self-energies are shown in \figdisp{sigma3D} for comparison. 

%\floatbarrier

In \figdisp{sigma3D}, the two theories have a similar pattern of k dependence, a dominant ridge running from left to right, and a less prominent feature running from top-left to bottom-right. They pass through $k=k_F, \omega=0$ region. The ridge leads to the appearance of twin peaks in the spectral functions representing spin-charge separation. In the higher energy region in \figdisp{sigma}, both theories agree well and are similar to their higher dimensional counterparts.

%\FloatBarrier

A powerful feature of  ECFL theory is that it allows us to vary temperature without extra effort,  at least in  the low to intermediate temperature region. In \figdisp{sigmaTscaling}, 
%the self-energy
 $\rho_{\Sigma}$ at $k_F$ is presented in several temperatures. The bump becomes higher with increasing temperature though no obvious change in larger scale (Panel (b)). This is expected because warming softens the peak height of spectral function at $k_F$, which is $\rho_G(k_F,0)=1/(\pi^2\rho_{\Sigma}(k_F,0)$ in Panel (c). The central peak  height $\rho_{\Sigma}(k_F,0)$ scales as $T^{\alpha}$ with $\alpha\approx1.1$,  as opposed to $\alpha=2$ expected for a Fermi liquid.
%qualitatively different from the typical $\alpha=2$ for Fermi liquid in higher dimensions.
 Although $T=0.005$ is the lowest temperature in the current numerical scheme for second order ECFL due to the finite lattice size (up to $L=2417$ and $N_\omega=2^{17}$), we extrapolate the curve to $T=0$.
 %plot the extrapolated $T=0$ curve, based on low temperature results. 
 The peak at $k_F$  disappears at zero T, and is replaced by a minimum at the origin corresponding to a singular peak in the spectral function, consistent with earlier studies \cite{1dhubbard, BA4}. The self-energy approaches zero as $\big|\omega\big|^\gamma$, where $\gamma\approx 1.3$. This behavior is difficult to observe in our present tDMRG implementation, because the finite time cut-off, leads to a broadening.  The peak and its $k$ dependence is recovered on moving  away from $k_F$, causing spin-charge separated peaks at T=0.
 %the zero temperature limit.

\begin{figure*}
\subfigure[\;\; ECFL T=0.005, J=0.3]{\includegraphics[width=.72\columnwidth]{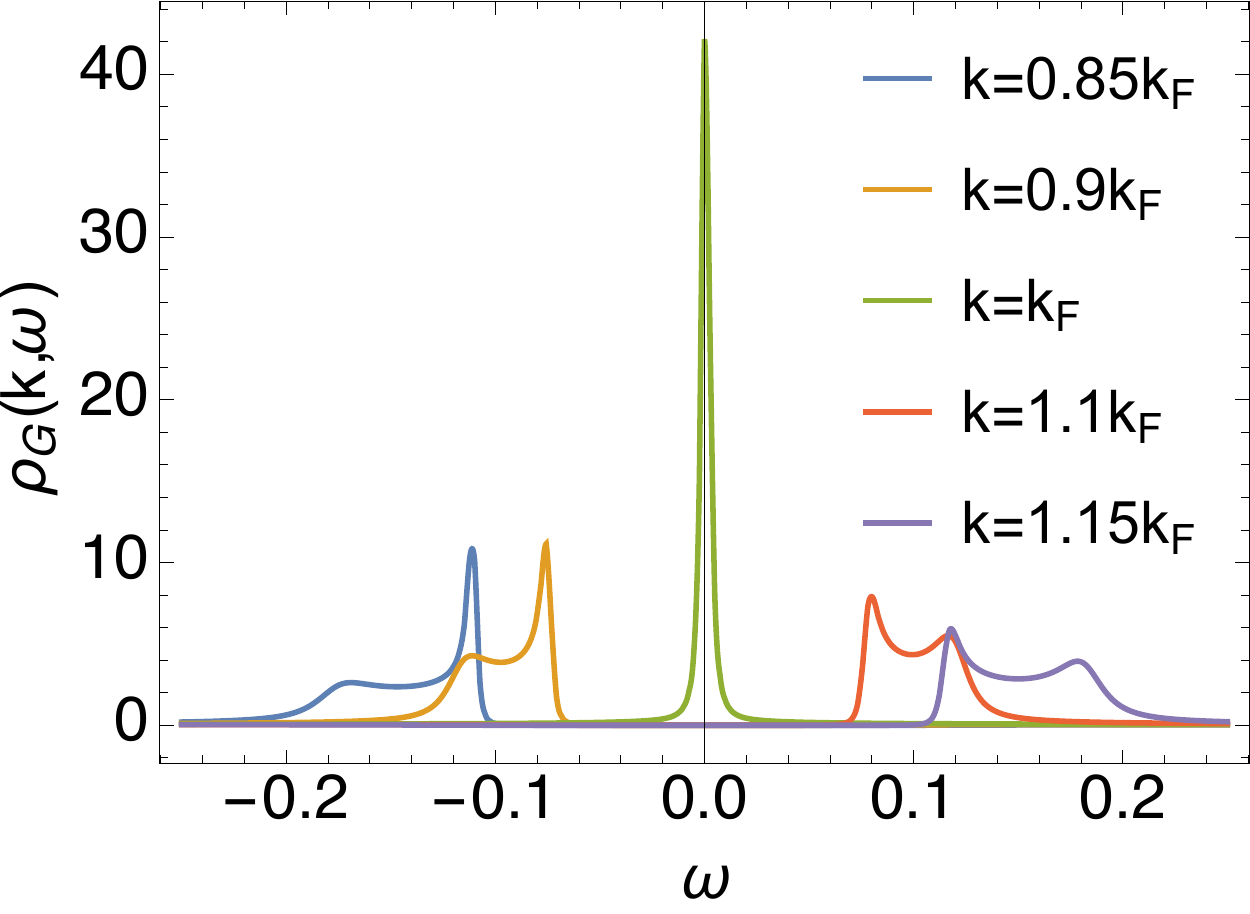}}
\subfigure[\;\; tDMRG T=0, J=0.3]{\includegraphics[width=.7\columnwidth]{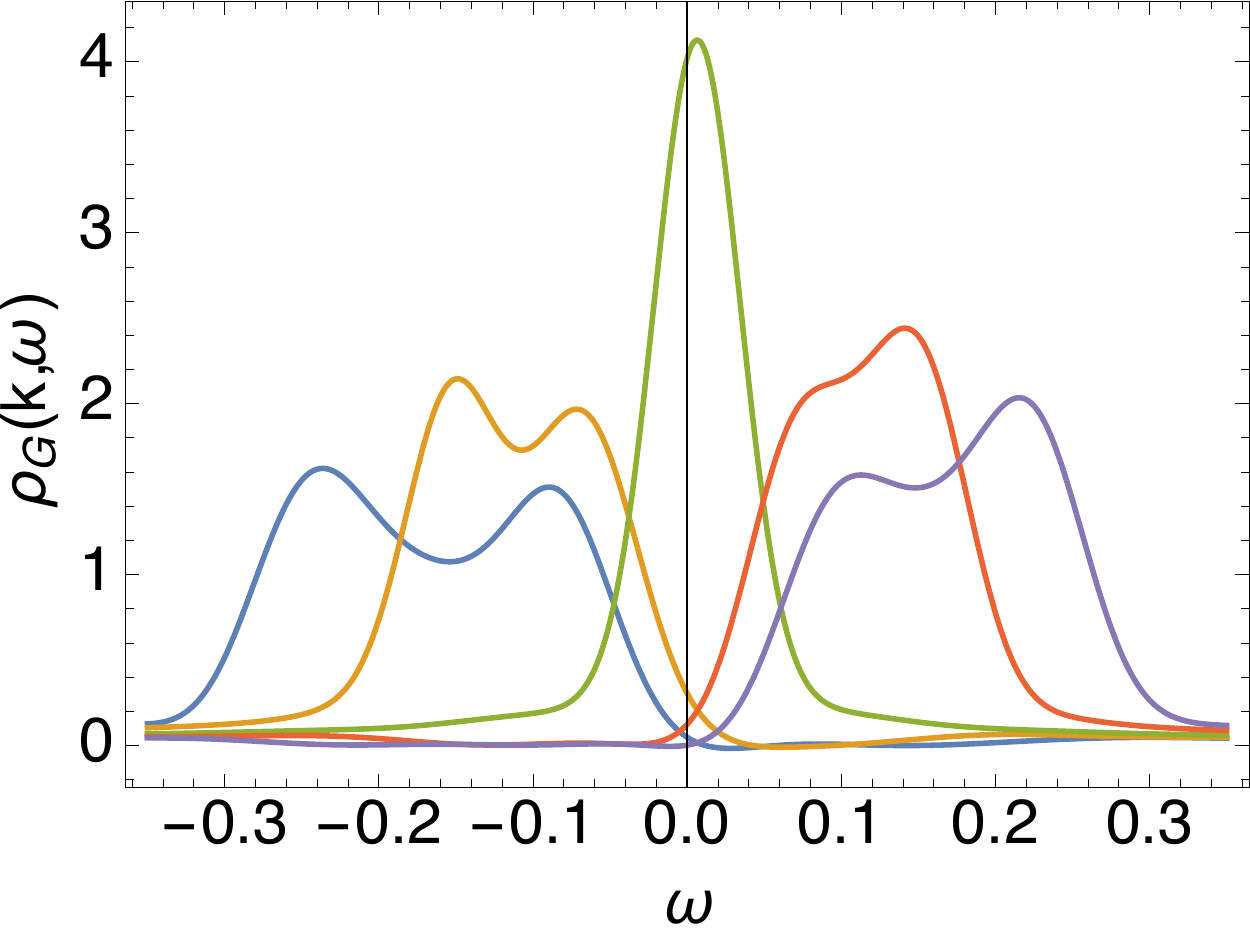}}
\subfigure[\;\; ECFL T=0.005, J=0.3]{\includegraphics[width=.73\columnwidth]{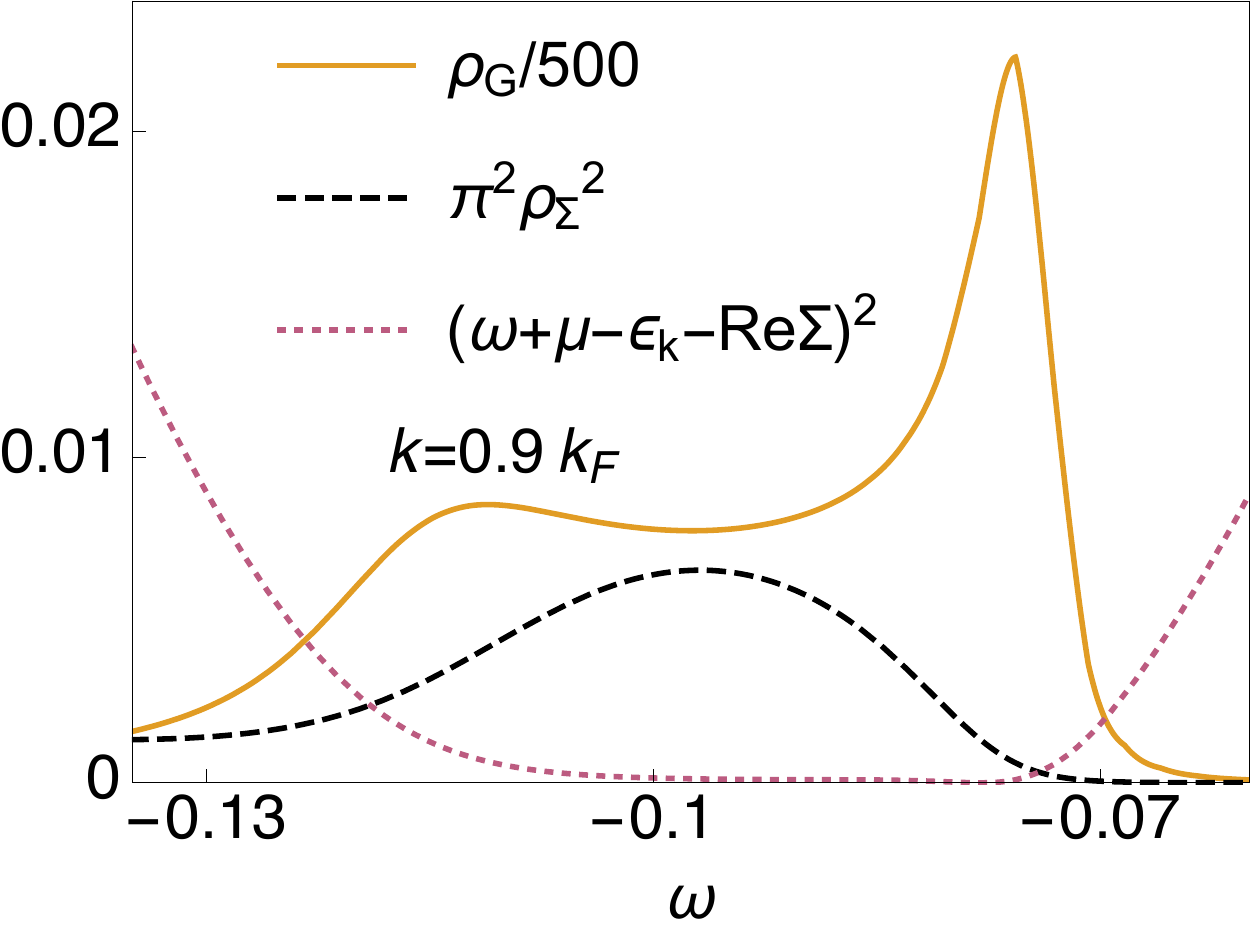}}
\subfigure[\;\; ECFL T=0.005, J=0.3]{\includegraphics[width=.7\columnwidth]{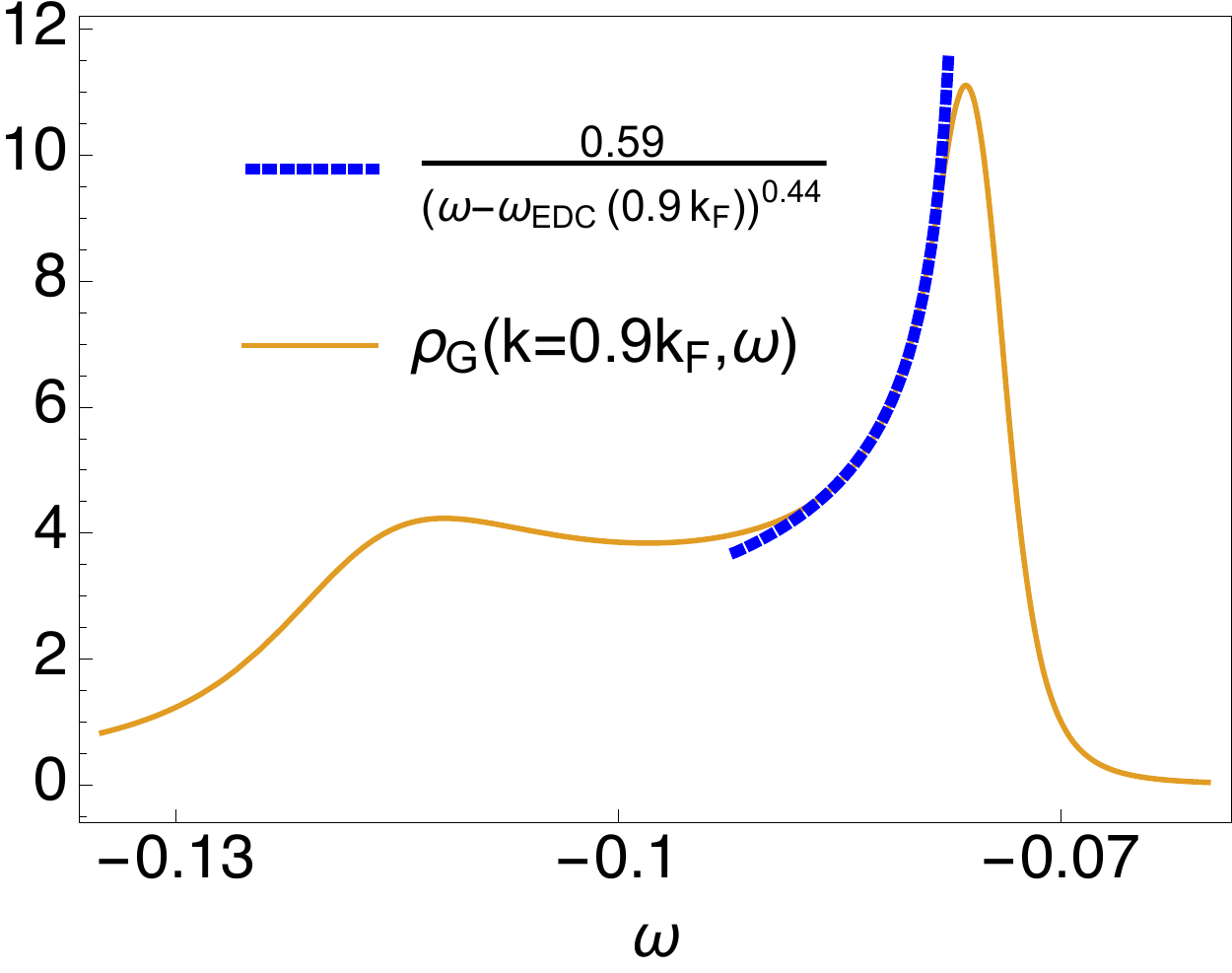}}
\caption{\label{fig5} Energy distribution curves (EDCs) at t'=0, J=0.3: (a) and (b) (same legends marking $k/k_F$) displaying the spinon and the holon for $k \neq k_F$. Panel (c) at $k=.9 k_F$ shows that the peak in $(\pi \rho_{\Sigma})^2$ (dashed black) coincides with the dip in the spectral function $\rho_G(\omega)$ (solid gold), while $(\omega+\mu-\varepsilon_k - \Re \, \Sigma)^2)$ (magenta dots) is small everywhere. This implies that the twin peaks originate in the intervening peak of self-energy.  
Panel (d) also at $k=.9 k_F$ shows the  fitting procedure for finding  the anomalous exponent $\zeta'\equiv \zeta-\frac{1}{2}$ for the spinon \cite{Giamarchi,Meden}, we fit to $.59(\omega- \omega_{peak})^{\zeta'}$ (dashed blue),
the best fit value is $\zeta' \sim -0.44$, close to the TLL result  $-0.45$ \cite{DMRG2}.
 }
\end{figure*}

\begin{figure*}
\subfigure[\;\; tDMRG]{\includegraphics[width=.95\columnwidth]{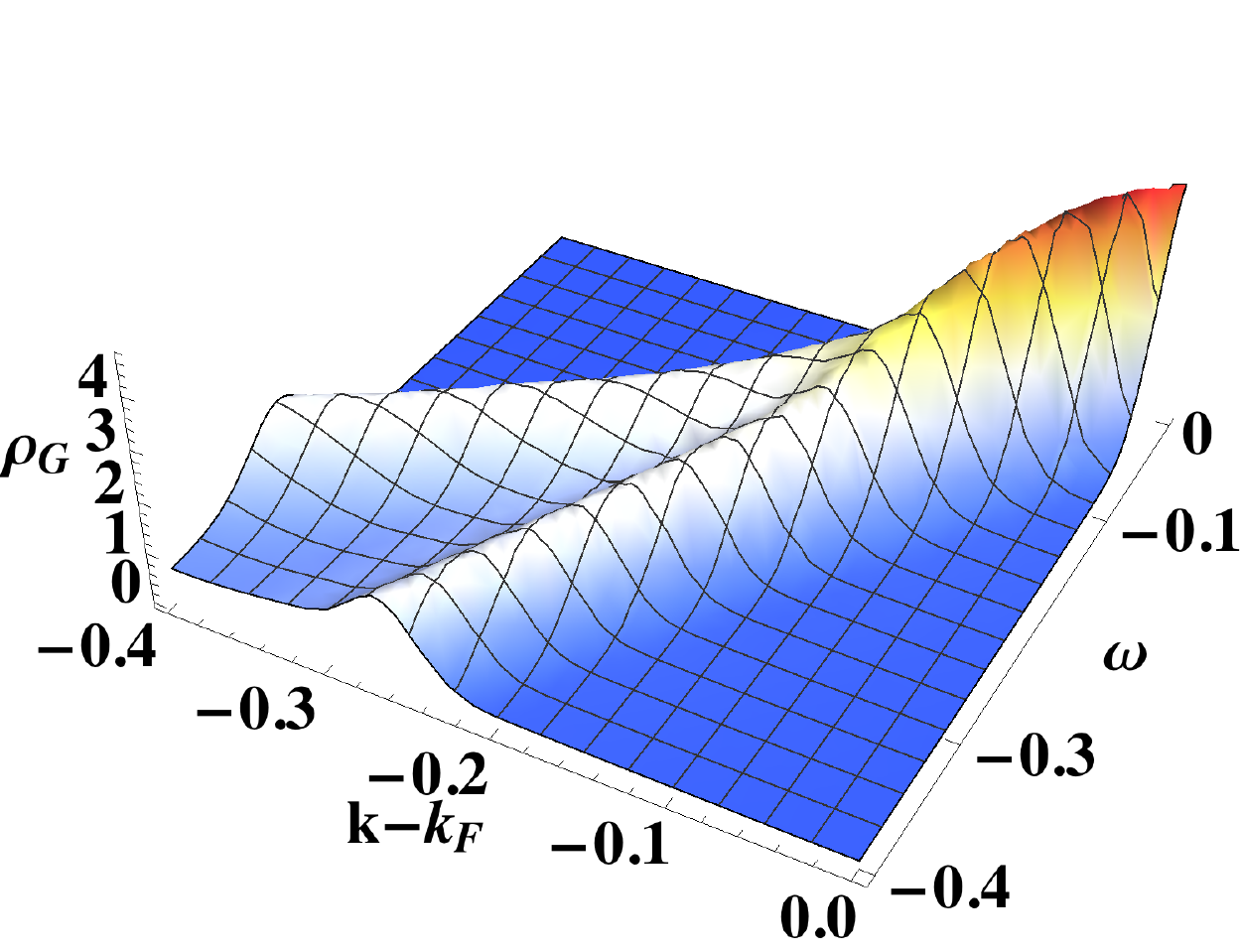}}
\subfigure[\;\; ECFL  with window]{\includegraphics[width=.95\columnwidth]{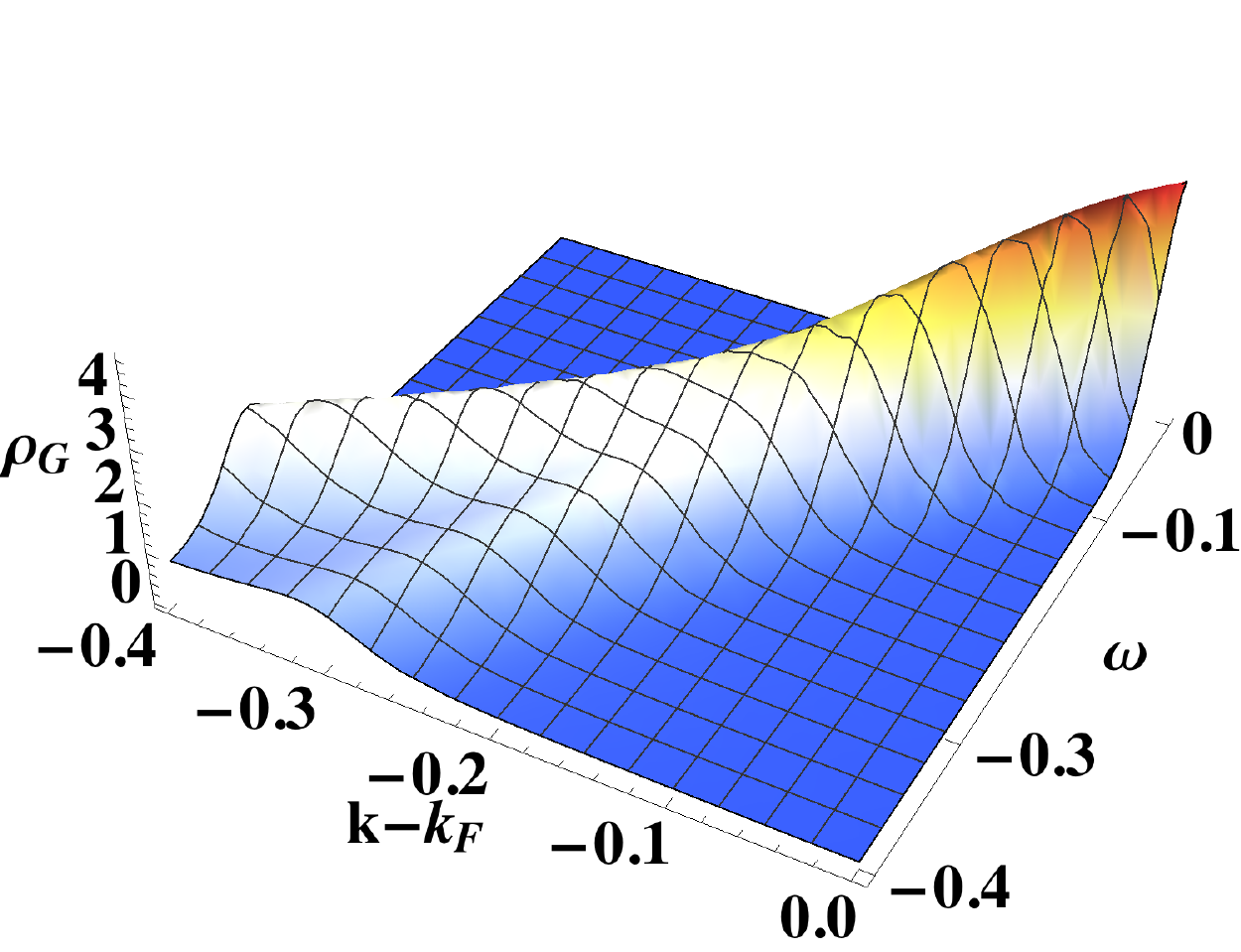}}
\subfigure[\;\; ECFL  without window]{\includegraphics[width=.95\columnwidth]{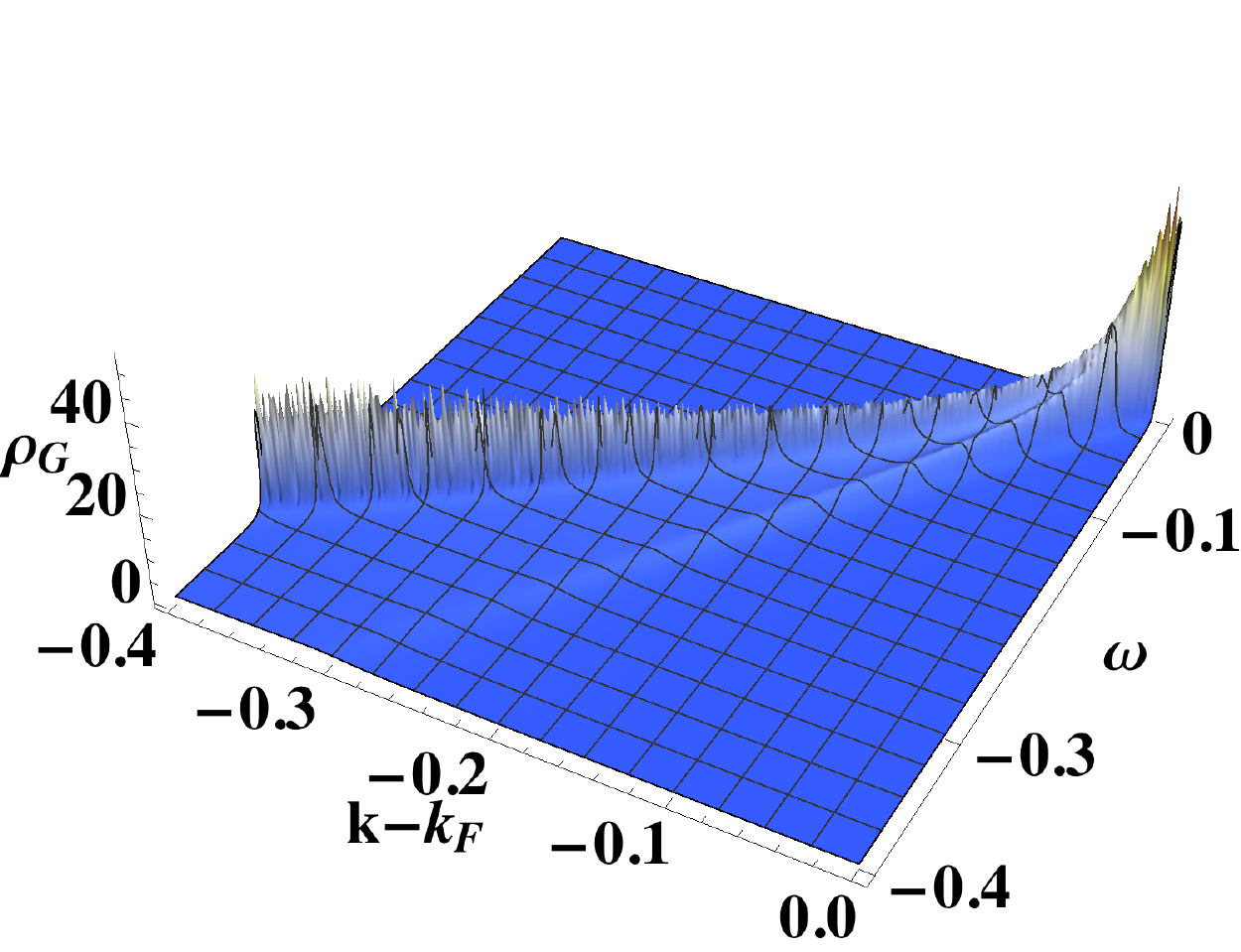}}
\caption{\label{EDC}  $J=0.6, t'=0$. The   spectral function of the  tDMRG ($T=0$)  with an intrinsic time window (a) and the ECFL ($T=.005$)  with (b) and without (c) a comparable  time window. The introduction of a time  window  brings the two theories to the same scale. The central peak and the spinon peaks  are of comparable height while the holon peak of ECFL is less prominent duo to second order approximation.
}
\end{figure*}
%\FloatBarrier

\begin{figure*}
\centering
\subfigure[\;\; n=0.7, t'=0, J=0.3]{\includegraphics[width=.7\columnwidth]{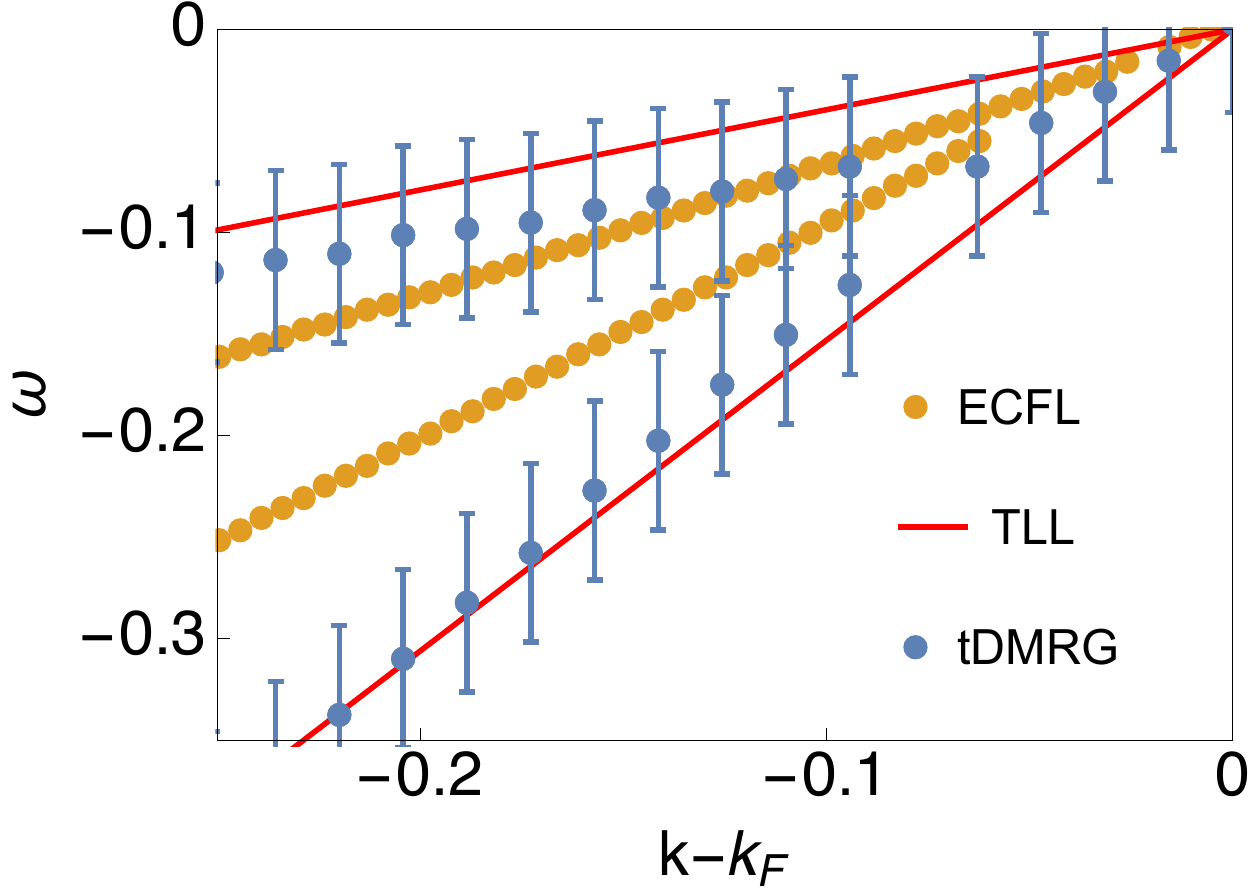}}
\subfigure[\;\; n=0.7, t'=0, J=0.6]{\includegraphics[width=.7\columnwidth]{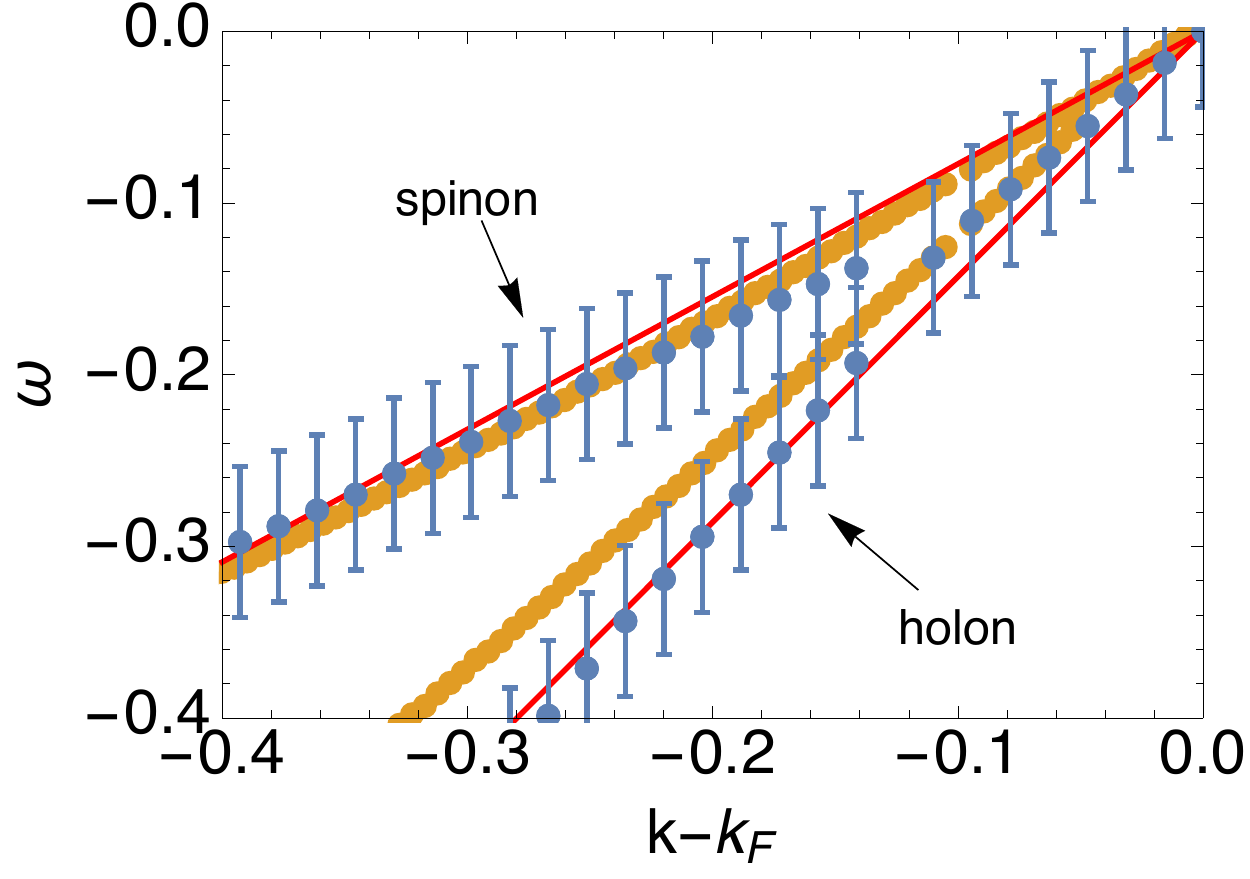}}
\subfigure[\;\; n=0.7, t'=0.2, J=0.3]{\includegraphics[width=.7\columnwidth]{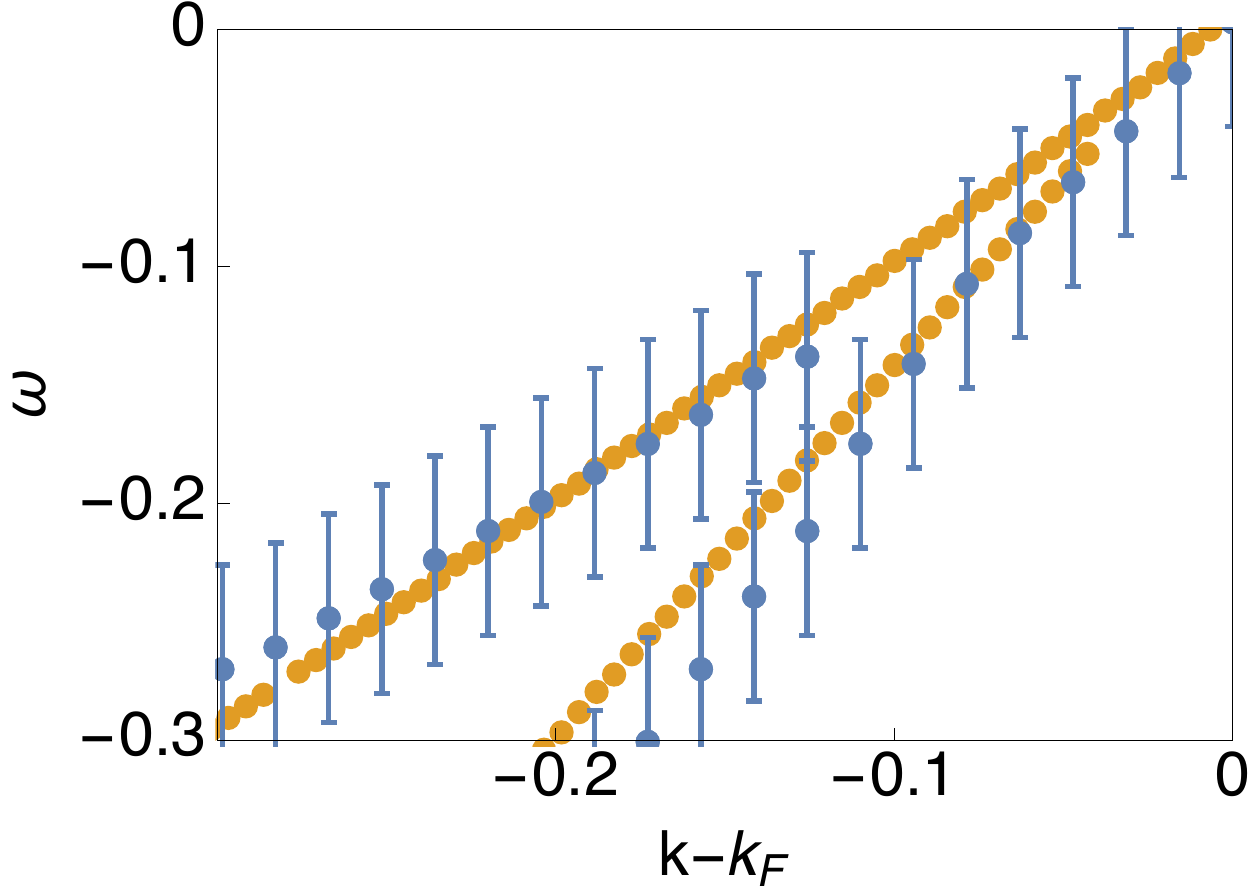}}
\subfigure[\;\; n=0.7, t'=0.2, J=0.6]{\includegraphics[width=.7\columnwidth]{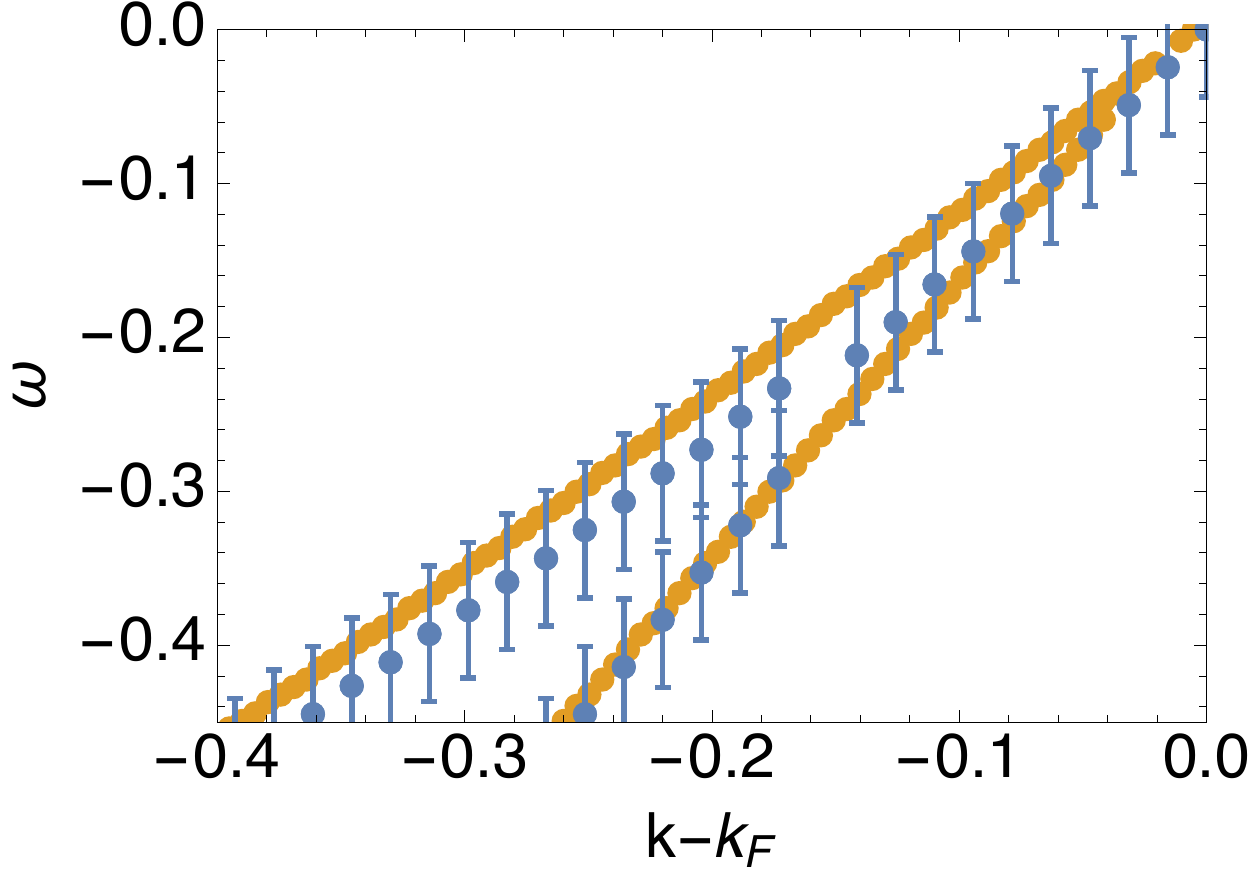}}
\caption{ \label{spectra}  Dispersion of excitations  from both ECFL at T=0.005 (gold dots) and tDMRG at T=0 (blue dots), and the available ED data (red) \cite{ed}.
% with n=0.7, J=0.3,0.6 and t'=0,0.2. 
The error bars in the tDMRG estimates are from the time window broadening.  The tDMRG results are consistent with the ED results, while the ECFL holon dispersion deviates somewhat.}
\end{figure*}

\section{\bf Spectral function}
We also compare the spectral functions from both methods. In \figdisp{fig5} panel (a,b) both show a single peak at $k_F$ and double peaks away from $k_F$ representing spinons and holons respectively. Panel (c) puts together the spectral function away from $k_F$ and different parts of its formula:
\beq
\rho_G(k,\omega)= \frac{\rho_\Sigma(k,\omega)}{[\omega+ \mu - \varepsilon_k -  \Re  \Sigma(k, \omega)]^2+ \pi^2 \rho^2_\Sigma(k,\omega)}, \label{formula}
\eeq
It shows that $\omega+ \mu - \varepsilon_k -  \Re  \Sigma(k, \omega) $ is very small in the frequency range that spans the two peaks, and confirms that the visible twin peaks result from a peak in $\rho_\Sigma$ in the middle.
Thus the location of the ridge lies in the minimum  between spinon and and holon peaks in the spectral function in panels (a,b), and in-fact the ridge causes the twin peaks.
 The exponents in panel (d) match reasonably with those from the TLL  at $J=0.3$  and also  at $0.6$ (where $\zeta'\sim -.49$ versus  $\zeta'\sim -0.46$ from  \refdisp{DMRG2}). We take the Luttinger parameter $K_{\rho}\approx 0.53$ at $J=0.3, t'=0$ from Fig. (4) in \refdisp{DMRG2}. Then we calculate $\zeta=\gamma_{\rho}=(K_{\rho}+K_{\rho}^{-1}-2)/8\approx 0.05$ \cite{Giamarchi,Meden}. Therefore the anomalous exponent is $\zeta'= \zeta-\frac{1}{2}=-0.45$. The calculation is similar for $J=0.6$ with $K_{\rho}\approx 0.56$ from Fig. (4). The tDMRG spectral function in panel b is too soft to extract the anomalous exponent, because its finite time cutoff leads to the broadening of spectral peaks in the low $\omega$ region. 

In \figdisp{EDC} we compares the spectral function of the  tDMRG with the ECFL theory. The latter is presented both with and without Gaussian windowing by a suitable time constant comparable to that in our  tDMRG work. As one might expect,  the scales of the two theories differ if we compare the raw (un-windowed) figures, but become very close upon windowing.

%in The ECFL spectral function also agrees with the asymtotic bosonization result with fitted parameters in the right inset of (a).  Panels (c) and (d) provide 3-d plots to view the spin-charge separation by showing two branches crossing at $k_F$.

\section{\bf Dispersion relation of spinons and holons}
%To study the spin-charge separation more explicitly, 
We extract the excitation dispersion relation from spectral function in \figdisp{spectra}. According to \refdisp{ed}, in the selected parameter region n=0.7, J=0.3, 0.6 and t'=0, 0.2, the holon velocity $v_c$ is larger than the spinon velocity $v_s$.  The error bars in the  tDMRG originates from the broadening of the lines due to  finite time windowing. Within the error bar, the DMRG agrees with the available ED data  \cite{ed}. We expect that the neglected higher order terms in the ECFL theory  would play a role in improving the holon velocity and also intensities.
%\FloatBarrier

\section {\bf Conclusion and Discussion} 
In this paper, we present the self-energy for the 1-d $t$-$t'$-$J$ model from both ECFL and tDMRG and specify its characteristic low energy strongly momentum-dependent cross-ridge, qualitatively different from higher dimensional cases,  responsible for  the spin-charge separation in spectral function. This perspective is different from the ones  discussed in earlier studies on this model  in 1-d \cite{CI,BA1,BA2,BA3,BA4,BA5,BA6,ed,DMRG1,DMRG2,QMC}. The existence of a ridge structure in the imaginary self-energy,  represents a non-trivial  exact statement about the  momentum dependence of the 1-d model.

We also compare the spectral function, the excitation dispersion and the momentum distribution function between both methods. They agree qualitatively in the  low energy region, both capturing clear signatures of the TLL and more quantitatively at larger energy scales where the system behaves like it does in higher dimensions.

  In summary we have  shown in this work  that the ECFL equations  capture the essential  physics of 1-d systems, namely  spin-charge separation and  non-Fermi liquid Green's functions in parallel to the  behavior displayed by the tDMRG solution. A  remarkable conclusion of this work  is that ECFL theory   works in the widely different regimes of infinite dimensions \cite{Sriram-Edward}, two dimensions \cite{SP,PS} and 1-d. This observation lends support to the overall scheme  in general  dimensions as well.

\section {\bf Acknowledgement} We thank  Rok \v{Z}itko for helpful  comments on the manuscript.
The work at UCSC was supported by the US Department of Energy (DOE), Office of Science, Basic Energy Sciences (BES), under Award No. DE-FG02-06ER46319. The work at UCI was supported by National Science Foundation (NSF) grant DMR-1505406. The ECFL Computations used the XSEDE Environment \cite{xsede} (TG-DMR170044) supported by National Science Foundation grant number ACI-1053575.

\end{document}